\documentclass[review]{elsarticle}

\usepackage[
     left=3cm,
    right=3cm
  ]{geometry}
\usepackage[utf8]{inputenc}
\usepackage[T1]{fontenc}
\usepackage{lmodern}
\usepackage{CJKutf8}
\usepackage[dvipsnames]{xcolor}
\usepackage{lineno,hyperref,amsmath,amsfonts}
\usepackage{cleveref}
\crefname{figure}{Fig.}{Figs.}
\crefname{table}{Tab.}{Tabs.}
\crefname{section}{Sec.}{Secs.}
\Crefname{figure}{Figure}{Figures}
\Crefname{table}{Table}{Tables}
\Crefname{section}{Section}{Sections}

\usepackage{tikz}
\usepackage{pgfplots}
\usepackage{grffile}
\pgfplotsset{compat=newest}
\hypersetup{
    colorlinks=true
}

\newcommand{\red}[1]{\textcolor{red}{#1}}
\newcommand{\blu}[1]{\textcolor{blue}{#1}}

\newcommand{\myi}[1]{\textcolor{MidnightBlue}{#1}}
\newcommand{\ket}[1]{\left|{#1} \right\rangle}

\newcommand{\braket}[2]{\left\langle {#1}\middle| {#2}\right\rangle}
\newcommand{\matX}{\mathbf{X}}
\newcommand{\matM}{\mathbf{M}}
\newcommand{\matP}{\mathbf{P}}
\newcommand{\matT}{\mathbf{T}}
\newcommand{\matL}{\mathbf{L}}
\newcommand{\matI}{\mathbf{I}}

\newcommand{\matV}{\mathbf{V}}
\newcommand{\matE}{\mathbf{E}}
\newcommand{\matB}{\mathbf{B}}
\newcommand{\matC}{\mathbf{C}}
\newcommand{\vecalpha}{\boldsymbol{\alpha}}
\newcommand{\vece}{\mathbf{e}}
\newcommand{\vecr}{\mathbf{r}}
\newcommand{\PfX}{\mathrm{Pf}\, \mathbf{X}}
\newcommand{\PfT}{\mathrm{Pf}\, \mathbf{T}}
\newcommand{\PfC}{\mathrm{Pf}\, \mathbf{C}}
\newcommand{\InvX}{\mathbf{X}^{-1}}
\newcommand{\InvT}{\mathbf{T}^{-1}}
\newcommand{\InvL}{\mathbf{L}^{-1}}
\newcommand{\InvC}{\mathbf{C}^{-1}}

\usetikzlibrary{arrows.meta,
                calc, chains,
                quotes,
                positioning,
                shapes.geometric,
                plotmarks}
\usepgfplotslibrary{patchplots}
\tikzstyle{Y1}=[fill=yellow, draw=black, shape=rectangle]
\tikzstyle{C1}=[fill=cyan, draw=black, shape=rectangle]
\tikzstyle{M1}=[fill={rgb,255: red,255; green,191; blue,191}, draw=black, shape=rectangle]

\journal{%Journal of \LaTeX\ Templates
         Computater Physics Communications}

\begin{document}

\begin{frontmatter}

\title{%Elsevier \LaTeX\ template\tnoteref{mytitlenote}
       Optimized Implementation for Calculation and Fast-Update of Pfaffians % \\
       Installed to the Open-Source Fermionic Variational Solver mVMC
       }
\tnotetext[mytitlenote]{%Fully documented templates are available in the elsarticle package on \href{http://www.ctan.org/tex-archive/macros/latex/contrib/elsarticle}{CTAN}.
                        Intro for software package \href{https://github.com/xrq-phys/Pfaffine}{Pfaffine} and its installation to
                        \href{https://github.com/xrq-phys/mVMC/tree/pfaffine-blocked}{mVMC} with additional fast-update techniques.}


\author[hongo]{RuQing G. Xu}
\ead{ruqing.xu@phys.s.u-tokyo.ac.jp}

\author[ipi,jst]{Tsuyoshi Okubo}
\author[hongo,ipi,issp]{Synge Todo} %\corref{Todo-lab}
% \cortext[Todo-lab]{\ead{wistaria@s.u-tokyo.ac.jp}}
\author[waseda,toyota]{Masatoshi Imada}

\address[hongo]{Department of Physics, University of Tokyo,
    Hongo 7-3-1, Bunkyo,
    Tokyo, 113-0033, Japan}
\address[ipi]{Institute for Physics of Intelligence, University of Tokyo,
    Hongo 7-3-1, Bunkyo,
    Tokyo, 113-0033, Japan}
\address[jst]{JST, PRESTO,
    Honcho 4-1-8, Kawaguchi,
    Saitama, 332-0012, Japan}
\address[issp]{Institute for Solid State Physics, University of Tokyo,
    Kashiwanoha 5-1-5, Kashiwa,
    Chiba,277-8581, Japan}
\address[waseda]{Deptartment of Applied Physics, Waseda University,
    Okubo 3-4-1, Shinjuku,
    Tokyo, 169-8555, Japan}
\address[toyota]{Toyota RIKEN, 
    Yokomichi 41-1, Nagakute,
    Aichi, 480-1118, Japan}

\begin{abstract}
In this article, we present a high performance, portable and well templated implementation for computing and fast-updating Pfaffian and inverse of an even-ranked skew-symmetric (antisymmetric) matrix.
It is achieved with a skew-symmetric, blocked variant of the Parlett-Reid algorithm and a blocked update scheme based on the Woodbury matrix identity.
Installation of this framework into the geminal-wavefunction-based \emph{many-variable Variational Monte Carlo} (mVMC) code boosts sampling performance to up to more than $6$ times without changing Markov chain's behavior. The implementation is based on an extension of the \emph{BLAS-like instantiation software} (BLIS) framework which has optimized kernel for many state-of-the-art processors including Intel Skylake-X, AMD EPYC Rome and Fujitsu A64FX.
\end{abstract}


\end{frontmatter}

% \linenumbers

% Suppress "Submitted to ..."
\thispagestyle{empty}

\section{Program Summary}
\label{sec:progsum}
\begin{tabular}{lp{9.5cm}}
  \textit{Program title:               } & Pfaffine and PfUpdates library for mVMC\cite{mVMC2019} \\
  \textit{CPC Library link to program: } & TBA  \\
  \textit{Developer's repository link:} & \url{https://github.com/issp-center-dev/mVMC/tree/master/src/pfupdates} \\
  \textit{                             } & \url{https://github.com/xrq-phys/Pfaffine} \\
  \textit{Code Ocean capsule:          } & TBA \\
  \textit{Licensing provisions:        } & MPL-2.0 (for new Library part) \\
  \textit{Programming language:        } & C++14 (for new Library part) \\
  % Library author: RuQing G. Xu \\
  % Supplementary material (if any): \\
  % Journal Reference of previous version:** \\
  % Does the new version supersede the previous version?:** \\
  % Reasons for the new version:** \\
  % Summary of revisions:** \\
  \textit{Nature of problem:           } & Finding a method for computing and updating Pfaffian and inverse of a skew-symmetric matrix that yields a high performance on modern processor architectures. \\
  \textit{Solution method:             } & Deploying a blocked version of the Parlett-Reid algorithm with BLIS serving as assembly-level backend. Updating is approached using a modified Woodbury matrix identity. \\
  % Additional comments including Restrictions and Unusual features (approx. 50-250 words): \\
  % References: \\
\end{tabular}

\section{Introduction}
\label{sec:intro}
Numerical variational methods for correlated many-body systems have played an important role in studies of novel condensed-matter phenomena such as high-temperature superconductivity and spin liquids\cite{white1993,sorella2002,clark2011,iqbal2013,misawa2014,hu2015,leblanc2015,ohgoe2020}.
% \begin{CJK}{UTF8}{ipxm}
    % \red{ここ Citation ちょっと増やすほうがいい}
    % Sorella / Imada 方法以外あんまり知らない故適当にググった
% \end{CJK}
Compared to exponentially expensive methods like exact diagonalization\cite{HPhi2017}, variational methods retains advantage that it demands polynomial computational complexity, yet is able to provide results of decent accuracy.
For Fermionic systems, there are several popular formulations of variational approaches. Density matrix renormalization group (DMRG)\cite{white1993}, which later evolved into the matrix product state (MPS)\cite{schollwock2011} ansatz, has achieved tremendous success in one-dimensional systems but suffers from lack of expressiveness in higher dimensions, while its extension for two-dimensional systems called projected entangled pair state (PEPS)\cite{verstraete2004,verstraete2008,maeshima2001} proved to be very challenging to optimize due to local minima problems\cite{corboz2016,vanderstraeten2016,liao2019}. On the contrary, the variational Monte Carlo method is easy to evaluate and optimize, while still capable to express those novel quantum states well\cite{tahara2008,morita2015,ido2018,nomura2017,nomura2020}.

During the past few years, several packages for variational Monte Carlo have been developed and opened to the public\cite{casino2000,nakano2020}, among them is mVMC\cite{mVMC2019} which focuses on the study of lattice-based systems. In mVMC, the variational wavefunction with $N_\mathrm{pair}$ spin-up Fermions and the same number of spin-down Fermions is taken to be the form:
\begin{equation}
    \label{eq:mvmc_ansatz}
    \ket \Psi = \mathcal L_s \mathcal L_p \mathcal P_J \mathcal P_G \ket \phi.
\end{equation}
In \cref{eq:mvmc_ansatz}, the core part $\ket \phi$ is given by:
\begin{equation}
    \ket \phi = \left( \sum_{i<j} f_{ij} c_i^\dag c_j^\dag \right)^{N_\mathrm{pair}} \left|0 \right\rangle,
\end{equation}
usually referred as a ``pair product'' (PP) or ``geminal wavefunction'' with $c_i^\dag$ as the Fermion creation operator on the $i$th spin-orbital, while $\mathcal P_J, \mathcal P_G$ are real-space correlation operators with Jastrow and Gutzwiller factors, respectively and $\mathcal L_s, \mathcal L_p$ are quantum-number projection operators for spin and momentum, respectively, to restore the original symmetry of the Hamiltonian. Readers interested in $\mathcal P_J, \mathcal P_G, \mathcal L_s$ and $\mathcal L_p$ are advised to refer to \cite{tahara2008} and the rest sections of this article only focus on PP.
With mean-field based $\ket \phi$\cite{tahara2008} providing a very good initial state and $\mathcal P_J, \mathcal P_G$ further considering effects introduced by interactions like Coulomb repulsion, mVMC is able to grab important features of many-body quantum mechanical systems\cite{morita2015,ido2018}, providing excellent approximations to their ground states as well as low-energy excited states \cite{ido2020,charlebois2020}.

Within the process of calculation, mVMC samples real space configurations $\left\{ \ket{x^{(k)}} \right\}$ using a Markov chain (MC) process. The probability for some configuration $\ket{x}$ to be sampled is proportional to its overlap with the wavefunction ansatz, i.e.:
\begin{equation}
    \label{eq:mvmc_mc_0}
    \mathrm{Prob}\left( \ket{x} \textrm{ is sampled} \right) = \frac{{\braket{x}\Psi}^2}{\braket \Psi \Psi}.
\end{equation}
According to the detailed balance condition of the Markov chain, \cref{eq:mvmc_mc_0} yields a stochastic process dominated by:
\begin{equation}
    \label{eq:mvmc_mc_1}
    \mathrm{Prob}\left( \ket{x} \longrightarrow \ket{x'} \right) =
        \min\left( 1, \frac{{\braket{x'}\Psi}^2}{{\braket{x}{\Psi}^2}} \right).
\end{equation}
As $\mathcal P_J$ and $\mathcal P_G$ are diagonal in configuration representation, while $\mathcal L_p$ and $\mathcal L_s$ expands to summed spatial operators, $\braket x \Psi$ expands to:
\begin{equation}
    \label{eq:mvmc_mc_ip_exp}
    \braket x \Psi = \sum_\theta \beta_\theta \sum_\vecr \alpha_\vecr \braket{T_\vecr(x)}{R_\theta(\phi)}.
\end{equation}
Here $\left| R_\theta (\phi) \right\rangle$ is $\left| \phi \right\rangle$ rotated by angle $\theta$ and $\left| T_\vecr(x) \right\rangle$ is $\left| x \right\rangle$ spatially translated by vector $\vecr$. Meshes for $(\theta, \vecr)$ and their corresponding weights are defined so as the original symmetry of Hamiltonian can be correctly restored. % to project the state onto specific spin- and momentum-quantum-numbers.
We recommended readers to refer to \cite{tahara2008,mVMC2019} for further details of those quantities.
\Cref{eq:mvmc_mc_ip_exp} tells us that to generate samples satisfying \cref{eq:mvmc_mc_0} we have to compute:
\begin{itemize}
    \item $\left\{\braket{T_r(x^{(0)})}{R_\theta(\phi)} \middle| r, \theta\right\}$ for the first sample;
    \item $\left\{\braket{T_r(x^{(k)})}{R_\theta(\phi)} \middle| r, \theta\right\}$ with
        $\left\{\braket{T_r(x^{(k-1)})}{R_\theta(\phi)} \middle| r, \theta\right\}$ known;
\end{itemize}
which are exactly the hotspots of the whole mVMC code.

Because $\ket{R_\theta(\phi)}$ generates a constant set of pair products $\ket{R_\theta(\phi)} = \left( \sum_{i<j} f_{ij}^\theta c_i^\dag c_j^\dag\right )^{N_\mathrm{pair}} \ket 0$ that does not change throughout the MC sampling process, and $\ket{T_r(x)}$ can be viewed as from another chain of samples with a constant displacement, we omit these transformation operators in later discussion and only use $\braket x \phi$ which is evaluated as:
\begin{equation}
    \label{eq:mvmc_ip_pfa}
    \braket x \phi = \PfX \textrm{ where } X_{\alpha \beta} = % F_{x_\alpha}^{x_\beta} =
    \left\{ \begin{array}{ll}
        0 & x_\alpha = x_\beta \\
        f_{x_\alpha x_\beta} & x_\alpha < x_\beta \\
        - f_{x_\beta x_\alpha} & \text{otherwise}
    \end{array} \right. ,
\end{equation}
where $x_\alpha$ denotes spin-orbital index of the $\alpha$-th Fermion under real-space configuration $\ket x$.
%\red{\begin{CJK}{UTF8}{ipxm}
%    ここ順番怪しい? Itemize の先に \cref{eq:mvmc_ip_pfa} を出すべき?
%    R_\theta, etc. の追記説明により、順番変更は不要になると思う。
%\end{CJK}}

Current implementation of mVMC\cite{mVMC2019} already features a nice threading behavior, making it able to run on large-scale computer clusters. However, this implementation bears defective microarchitecture usage on modern processor hardware due to the following two reasons:
\begin{itemize}
\item Heaviest calculation of the mVMC method involves operations on skew-symmetric (antisymmetric) matrices, which is not available via the standard BLAS interface. The alternative library, known as Pfapack77\cite{Wimmer2011}, is programmed without considering pipeline-level hardware efficiency hence not optimal for those state-of-the-art processors.
\item To avoid recalculating Pfaffian every time for each Markov chain update, a fast-update scheme is adopted in current mVMC implementation. However, only rank-1 update scheme is implemented. The rank-1 update strategy is, by its name, mainly invoking rank-1 updates of matrices $\left( \mathbf X = \mathbf X + \mathbf{u v}^T \right)$, which is doomed to hit the memory bound and impact performance.
\end{itemize}
To get around these two severe hurdles upon mVMC's performance, a new fast-update scheme as well as a new implementation of the skew-symmetric matrix library is necessary so that:
\begin{itemize}
\item All computational hotspots are effectively handled with assemblies fully optimized for the hosting hardware.
\item Fast-update of Pfaffian invokes matrix-matrix operations instead of rank-1 updates.
\end{itemize}
Such are exactly what our optimization does.

This article is organized in the following fashion: \Cref{sec:pfaffine} reviews and extends Wimmer's approach in Pfapack77\cite{Wimmer2011} on tridiagonalizing skew-symmetric matrices and computing Pfaffians. \Cref{sec:pfupdates} presents the blocked fast-update scheme for further speeding up Pfaffian updates.
\Cref{sec:api} specifies programming interfaces of this acceleration framework as well as our assembly-level tuned implementation.
\Cref{sec:benchmark} shows benchmarking results indicating superiority of this new implementation and finally, \cref{sec:conclusion} concludes our development work.

\section{Skew-symmetric Matrix and Pfaffian}
\label{sec:pfaffine}
Pfaffian of a size $2n \times 2n$ skew-symmetric matrix $\mathbf X$ is defined as:
\begin{equation}
    \label{eq:pfa}
    \PfX = \frac 1{2^n n!} \sum_{\sigma \in S_{2n}} \mathrm{sgn} (\sigma) \prod_i^n X_{\sigma(2i - 1), \sigma(2i)},
\end{equation}
and satisfies:
$$
    \left(\PfX \right)^2 = \det \matX,
$$
where $S_{2n}$ is the permutation group of sets with $2n$ elements. Directly evaluating \cref{eq:pfa} is of exponential complexity due to the number of elements in $S_{2n}$. However, there are linear transformations that can be applied to simplify computation just like the case of calculating determinants.

Before introducing tridiagonalization computation of Pfaffian, let us first go over definition of the Gaussian elimination matrix:
\begin{equation}
    \label{eq:sktdf_elim1c}
    \matM_k = \matI_{2n} - \vecalpha_k \vece_k^T,
\end{equation}
where $\matI_{2n}$ is the $2n\times 2n$ identity, $\vece_k$ is the $k$-th elementary vector of Euclidean space
(i.e. $\vece_k = (\underbrace{0, \cdots, 0}_{k-1 \text{ zeros}}, 1,
    \underbrace{0, \cdots, 0}_{2n-k \text{ zeros}})^T$)
and $\vecalpha_k$ is constructed with entries:
\begin{equation}
    \left(\alpha_k \right)_i = \left\{ \begin{array}{ll}
        0 & i \le k \\
        \left. X_{i, k-1} \middle/ X_{k, k-1} \right. & i > k
    \end{array} \right.
\end{equation}
so that left-applying $\matM_k$ to $\matX$ eliminates all
  $\left. X_{i, k-1} \right|_{i > k}$ with
  $\left. X_{ij} \right|_{i \le k, \forall j}$ untouched.
Applying $\matM_k$ to both sides of a skew-symmetric matrix $\matX$ eliminates non-tridiagonal part of the $\left( k-1 \right)$-th column and row without changing its Pfaffian\cite{Wimmer2011}. This will be the main weapon for efficient Pfaffian calculation in the formulation of this section.

\paragraph{Blocked Tridiagonal Decomposition}
Blocked tridiagonal decomposition of skew-symmetric matrices (\texttt{sktdf}) was first formulated by Wimmer in Pfapack77\cite{Wimmer2011}. In their work a blocked and skew-symmetric variant of the Parlett-Reid algorithm\cite{Parlett1970} is adopted so that:
\begin{align} \label{eq:sktdf}
    \matP \matX \matP^T &= \matL \matT \matL^T, \\
    \matP &= \matP_{n-1} \cdots \matP_{2}, \nonumber\\
    \matT &=
        \matM_{\left[ n-1-r^{(s)} :: n-1 \right]} \cdots \matM_{\left[ 2 :: 2+r^{(1)} \right]} \matX
        \matM_{\left[ 2 :: 2+r^{(1)} \right]}^T \cdots \matM_{\left[ n-1-r^{(s)} :: n-1 \right]}^T \nonumber\\
    \matL &=\left(
        \matM_{\left[ n-1-r^{(s)} :: n-1 \right]} \cdots
        \matM_{\left[ 2 :: 2+r^{(1)} \right]} \matP^{T}\right)^{-1} \nonumber \\
    \matM_{\left[ i :: i+r^{(t)} \right]} &=
        \matM_{i+r^{(t)}} \matP_{i+r^{(t)}} \cdots \matM_{i+1} \matP_{i+1} \matM_{i} \matP_{i}. \nonumber
\end{align}
Here $\matP_k$ are permutation matrices that swap the $k$-th and $\left( p(k)>k \right)$-th columns (rows) of $\matX$ when multiplied on the right (left) side.
We call linear transformation $\matM_{\left[ i :: i+r^{(t)} \right]}$ a rank-$r^{(t)}$ pivoted Gaussian elimination where $r^{(t)} \in \mathbb{Z}^+$ is a parameter for performance tuning. It is constructed by merging multiple Gaussian elimination matrices:
\begin{align}
\matM_{\left[k :: k+1\right]}
    &= \matM_{k+1} \matP_{k+1} \matM_k  \nonumber \\
    &= \left( \matI_{2n} - \vecalpha_{k+1} \vece_{k+1}^T \right) \matP_{k+1}
       \left( \matI_{2n} - \vecalpha_k \vece_k^T \right) \nonumber \\
    &= \left( \matI_{2n} -
    \underbrace{ \begin{bmatrix} \vecalpha_{k+1} & \vecalpha^{(p)}_k
        \end{bmatrix} }_{\matV_{\left[k :: k+1\right]}}
    \underbrace{ \begin{bmatrix}
        \vece_{k+1}^T \\
        \vece_k^T
    \end{bmatrix} }_{\matE_{\left[k :: k+1\right]}}
    \right) \matP_{k+1} \label{eq:sktdi_blk2c} \\
\matM_{\left[k-r :: k+1\right]}
    &= \matM_{k+1} \matP_{k+1} \matM_{\left[k-r :: k\right]} \nonumber \\
    &= \left( \matI_{2n} - \vecalpha_{k+1} \vece_{k+1}^T \right) \matP_{k+1}
       \left( \matI_{2n} - \matV_{\left[k-r :: k\right]}
                           \matE_{\left[k-r :: k\right]}^T \right)
       \matP_k \cdots \matP_{k-r} \nonumber \\
    &= \left( \matI_{2n} -
    \underbrace{ \begin{bmatrix} \vecalpha_{k+1} & \matV_{\left[k-r :: k\right]}^{(p)}
        \end{bmatrix} }_{\matV_{\left[k-r :: k+1\right]}}
    \underbrace{ \begin{bmatrix}
        \vece_{k+1}^T \\
        \matE_{\left[k-r :: k\right]}^T
    \end{bmatrix} }_{\matE_{\left[k-r :: k+1\right]}}
    \right) \matP_{k+1} \matP_k \cdots \matP_{k-r}, \label{eq:sktdi_blkrc}
\end{align}
where $\matV_{\left[... k\right]}^{(p)}$ and $\vecalpha^{(p)}_k$ denote a matrix and column vector with $k$-th and $p(k)$-th rows swapped. Note also that in derivation of \cref{eq:sktdi_blk2c,eq:sktdi_blkrc} we used $\vece_{k+1}^T \vecalpha_k = 0$ and $\vece_k = \matP_{k+r} \vece_k$ implied by \cref{eq:sktdf_elim1c}.
We refer to \cite{Wimmer2011} for further details of this algorithm.

Once factorization \eqref{eq:sktdf} is done, with the fact that neither $\matP_k$ or $\matM_k$ affects the matrix's Pfaffian, $\PfX$ can be computed as:
\begin{equation}
    \label{eq:skpfa}
    \PfX = \PfT = \prod_{i=1}^n T_{2i-1, 2i}.
\end{equation}

\paragraph{Inversion from Tridiagonalization}
With \cref{eq:skpfa} completes Wimmer's derivation in \cite{Wimmer2011}, but mVMC requires computation of $\InvX$ in addition to $\PfX$. Current version of the implementation is initializing another \texttt{GETRF}/\texttt{GETRI} call to invert $\matX$ as a non-symmetric matrix. This treatment requires $\sim (2n)^3$ multiply-accumulate operations (MACs). However, one might notice that by inverting \cref{eq:sktdf}, $\InvX$ can be expressed as:
\begin{align}
    \label{eq:ltl2inv}
    \matP \InvX \matP^T &= \left(\InvL \right)^T \InvT \InvL \\ &= \matP
        \matM_{\left[ 2 :: 2+r^{(1)} \right]}^T \cdots
        \matM_{\left[ n-1-r^{(s)} :: n-1 \right]}^T \matT^{-1}
        \matM_{\left[ n-1-r^{(s)} :: n-1 \right]} \cdots
        \matM_{\left[ 2 :: 2+r^{(1)} \right]} P^T \nonumber \\
    \label{eq:sktdi}
    \Rightarrow \InvX &=
        \matM_{\left[ 2 :: 2+r^{(1)} \right]}^T \cdots
        \matM_{\left[ n-1-r^{(s)} :: n-1 \right]}^T \matT^{-1}
        \matM_{\left[ n-1-r^{(s)} :: n-1 \right]} \cdots
        \matM_{\left[ 2 :: 2+r^{(1)} \right]},
\end{align}
so that one can either invert $\matX$ by expanding \cref{eq:sktdi} from the center or by directly evaluating \cref{eq:ltl2inv} since $\matL$ is obtainable by properly merging and swapping all $\matV$ entries of $\matM$\cite{Wimmer2011}.

\paragraph{Quantitative Cost of Computation}
Evaluation of $\matM_{\left[k-r :: k\right]} \matX^{(k-r-1)} \matM_{\left[k-r :: k\right]}^T$ can be done as:
\begin{align}
    \matM_{\left[k-r :: k\right]} \matX^{(k-r-1)} \matM_{\left[k-r :: k\right]}^T &=
        \left( \matI - \matV_{\left[k-r :: k\right]}
                       \matE_{\left[k-r :: k\right]}^T \right)
    \matX^{(k-r-1, p)}
    \left( \matI - \matV_{\left[k-r :: k\right]}
                   \matE_{\left[k-r :: k\right]}^T \right)^T \nonumber \\
  &= \matX^{(k-r-1, p)} +
    \matV_{\left[k-r :: k\right]} \matX^{(k-r-1, p) T}_{\left[k-r :: k\right], [::]} -
    \matX^{(k-r-1, p)}_{\left[k-r :: k\right], [::]} \matV_{\left[k-r :: k\right]}^T, \label{eq:sktdf_evalrc}
\end{align}
where we have used during the derivation that $\matX$ is skew-symmetric and $\matV$'s columns are orthogonal to each other. $\matX^{(k-r-1)}$ denotes $\matX$ after $k - r - 1$ steps of Gaussian elimination and $\matX^{(k-r-1, p)}$ its permuted version, i.e.\footnote{
    Here for ease notation, an unblock expression is used. In real computation we always adopt blocked manipulations.
}:
\begin{align*}
    \matX^{(k-r-1)} &= \matM_{k-r-1} \matP_{k-r-1} \cdots \matM_2 \matP_2
        \matX^{(k-r-1)} \matP_2^T \matM_2^T \cdots \matP_{k-r-1}^T \matM_{k-r-1}^T, \\
    \matX^{(k-r-1, p)} &= \matP_{k}\cdots \matP_{k-r}
        \matX^{(k-r-1)} \matP_{k-r}^T \cdots \matP_{k}^T.
\end{align*}
As $\matX^{(k-r-1, p)}$ always has the first $k-r-1$ columns and rows tridiagonal, evaluation of \cref{eq:sktdf_evalrc} only needs to consider the lower-right portion of size $\left(2n - (k-r-1)\right) \times \left(2n - (k-r-1)\right)$. This fact together with $\matX$'s skew-symmetric property yields an evaluation cost of $(2n - (k-r-1))^2 r$. Hence the total evaluation cost to transform $\matX$ to $\matT$ in \cref{eq:sktdf} is around $ \frac 1 3 (2n)^3$ MACs.

For extracting $\matX^{-1}$, instantiation of \cref{eq:sktdi} requires iteratively evaluating $\matM_{\left[k-r :: k\right]}^T \matT^{-1}_{(k-r-1)} \matM_{\left[k-r :: k\right]}$ which expands into:
\begin{align}
& \matM_{\left[k-r :: k\right]}^T \matT^{-1}_{(k-r-1)} \matM_{\left[k-r :: k\right]} \nonumber \\
  &=\matP_{k-r} \cdots \matP_k
    \left( \matI_{2n} - \matV_{\left[k-r :: k\right]}
                        \matE_{\left[k-r :: k\right]}^T \right)^T
    \matT_{(k-r-1)}^{-1}
    \left( \matI_{2n} - \matV_{\left[k-r :: k\right]}
                        \matE_{\left[k-r :: k\right]}^T \right) \matP_k \cdots \matP_{k-r} \nonumber \\
  &=\matP_{k-r} \cdots \matP_k \Big\{
        \matT_{(k-r-1)}^{-1} -
            \matE_{\left[k-r :: k\right]} \left( \matT_{(k-r-1)}^{-1} \matV_{\left[k-r :: k\right]} \right)^T
    \nonumber \\ &\quad +
            \left( \matT_{(k-r-1)}^{-1} \matV_{\left[k-r :: k\right]} \right) \matE_{\left[k-r :: k\right]} +
            \matE_{\left[k-r :: k\right]} \matV_{\left[k-r :: k\right]}^T \matT_{(k-r-1)}^{-1}
                                          \matV_{\left[k-r :: k\right]} \matE_{\left[k-r :: k\right]}
    \Big\} \matP_k \cdots \matP_{k-r}. \label{eq:sktdi_evalrc}
\end{align}
Computation of matrix-matrix products in \cref{eq:sktdi_evalrc} requires around $(2n) \left(2n - \left(k-r-1\right) \right) r + (2n) r^2$ MACs. Ignoring $(2n) r^2$ term which is much less than $(2n) \left(2n - \left(k-r-1\right) \right) r$, total cost of \cref{eq:sktdi} becomes $\approx \frac12 (2n)^3$.

The alternative approach for $\matX^{-1}$, namely \cref{eq:ltl2inv} requires inverting $\matL^{-1}$ first, calling a subprocedure of \texttt{GETRI} named \texttt{TRTRI} which makes $\frac16 (2n)^3$ MACs. Since $\matT$ is tridiagonal, computation of $\matT^{-1} \matL^{-1}$ can be done by using the $O\left(n^2 \right)$-cost procedure \texttt{GTSV} whose cost can be neglected. The final $\matX^{-1} = \left(\matL^{-1}\right)^T \left(\matT^{-1}\matL^{-1}\right)$ step has triangular $\matL^{-1}$ on the left hand side and only needs to write half of $\matX^{-1}$ since we already know that it is skew-symmetric, resulting in $\frac16 (2n)^3$ MACs. Hence the whole evaluation of \cref{eq:ltl2inv} totals up to $\frac13 (2n)^3$ MACs,
%% can be evaluated with \texttt{TRMM} since triangular matrix $\matL$'s inverse is also triangular, yielding $\frac12 (2n)^3$ MACs.
also lower than \texttt{GETRF}/\texttt{GETRI} deployed by the original algorithm. Both approaches for $\matX^{-1}$ are available in our implementation as different subversions.

\section{Blocked Pfaffian Update from Modified Woodbury Matrix Identity}
\label{sec:pfupdates}
\paragraph{Theory}
mVMC takes configuration samples $\left\{ \ket{x^{(k)}} \middle| k\in \mathbb N \right\}$ with a Markov chain (MC) process, as is shown in \cref{eq:mvmc_mc_1}. As each step $\ket{x^{(k)}} \rightarrow \ket{x^{(k+1)}}$ within this MC only changes the spin-orbitals of one or two electrons of $\ket{x^{(k)}}$, we do not need to redo \cref{sec:pfaffine} calculations each time a new sample is generated. Instead, from the Woodbury matrix identity\cite{woodbury1950} we have:
\begin{align}
    \mathrm{Pf}\left[\matX + \matB \matC \matB^T \right] &= \PfX \times
        \frac{\mathrm{Pf}\left[ \InvC + \matB^T \InvX \matB \right]}
            {\mathrm{Pf}\left[ \InvC \right]} \\
    \left(\matX + \matB \matC \matB^T \right)^{-1} &= \InvX -
        \InvX \matB \left(\InvC + \matB^T \InvX \matB \right)^{-1} \matB^T \InvX,
\end{align}
which allows Pfaffian in \cref{eq:mvmc_ip_pfa} to be updated as % \footnote{}
    (To give a more intuitive image of what is happening during the MC fast-update, we also plug in here a $4\times 4$ exemplar matrix for $\mathbf X$.):

\begin{align}
\ket x &= \ket{ {j_1} {j_2} {j_3} {j_4} } \\
\mathbf X &= \left[ \begin{array}{cccc}
  0             &  F_{j_1}^{j_2} &  F_{j_1}^{j_3} & F_{j_1}^{j_4} \\
 -F_{j_1}^{j_2} &  0             &  F_{j_2}^{j_3} & F_{j_2}^{j_4} \\
 -F_{j_1}^{j_3} & -F_{j_2}^{j_3} & 0              & F_{j_3}^{j_4} \\
 -F_{j_1}^{j_4} & -F_{j_2}^{j_4} & -F_{j_3}^{j_4} & 0
\end{array}\right] \\
\braket x \phi &= \PfX \\
%\mathbf X^{-1} &= \mathbf X^{-1} \\
 & \downarrow \nonumber \\
\ket {x^{(1)}} &= \ket{ \red{i_1} {j_2} {j_3} {j_4} } \\
\mathbf X^{(1)} &= \left[ \begin{array}{cccc}
        0              &  \red{F_{i_1}^{j_2}} &  \red{F_{i_1}^{j_3}} & \red{F_{i_1}^{j_4}} \\
  \red{-F_{i_1}^{j_2}} &       0              &       F_{j_2}^{j_3}  &      F_{j_2}^{j_4} \\
  \red{-F_{i_1}^{j_3}} &      -F_{j_2}^{j_3}  &       0              &      F_{j_3}^{j_4} \\
  \red{-F_{i_1}^{j_4}} &      -F_{j_2}^{j_4}  &      -F_{j_3}^{j_4}  &      0
\end{array}\right]
 = \mathbf X + \mathbf B^{(1)} \underbrace{\left[\begin{array}{cc}
  0 & -1 \\
  1 &  0
\end{array}\right]}_{\mathbf C^{(1)}} \mathbf{B^{(1)}}^T \\
B^{(1)}_{lm} &= \left\{ \begin{array}{ll}
  F_{i_1}^{j_l} - F_{j_1}^{j_l} & l \neq 1, m = 1 \\
  1                             & l =    1, m = 1 \\
  0                             & \textrm{otherwise}
  \end{array}\right. \\
\label{eq:pfu_pfa1}
\braket{x^{(1)}}{\phi} &= \PfX^{(1)} = \PfX \times \left.
 \mathrm{Pf} \left(\mathbf C^{(1)} + \mathbf{B^{(1)}}^T \mathbf X^{-1}\mathbf B^{(1)}\right) \right/
  \PfC^{(1)}
 = \PfX \times \left(1 - \sum_{j} \left(X^{-1}\right)_{1}^{j} u_j\right) \\
 \label{eq:pfu_inv1}
\mathbf{X^{(1)}}^{-1} &= \mathbf X^{-1} - \mathbf X^{-1} \mathbf B^{(1)}
 \left(\mathbf C^{(1)} + \mathbf{B^{(1)}}^T \mathbf X^{-1}\mathbf B^{(1)}\right)^{-1}
  \mathbf{B^{(1)}}^T \mathbf X^{-1} \\
 & \downarrow \nonumber
\end{align}

\begin{align}
\ket {x^{(2)}} &= \ket{ \red{i_1} \red{i_2} {j_3} {j_4} } \\
\mathbf X^{(2)} &= \left[ \begin{array}{cccc}
        0              &  \blu{F_{i_1}^{i_2}} &  \red{F_{i_1}^{j_3}} & \red{F_{i_1}^{j_4}} \\
  \blu{-F_{i_1}^{i_2}} &       0              &  \red{F_{i_2}^{j_3}} & \red{F_{i_2}^{j_4}} \\
  \red{-F_{i_1}^{j_3}} & \red{-F_{i_2}^{j_3}} &       0              &      F_{j_3}^{j_4} \\
  \red{-F_{i_1}^{j_4}} & \red{-F_{i_2}^{j_4}} &      -F_{j_3}^{j_4}  &      0
\end{array}\right]
 = \mathbf X + \mathbf B^{(2)} \mathbf C^{(2)} \mathbf{B^{(2)}}^T \\
 \label{eq:pfu_pfa2}
\braket{x^{(2)}}{\phi} &= \PfX^{(2)} = \PfX \times \left.
 \mathrm{Pf} \left(\mathbf C^{(2)} + \mathbf{B^{(2)}}^T \mathbf X^{-1}\mathbf B^{(2)}\right) \right/
  \PfC^{(2)} \\
 \label{eq:pfu_inv2}
\mathbf{X^{(2)}}^{-1} &= \mathbf X^{-1} - \mathbf X^{-1} \mathbf B^{(2)}
 \left(\mathbf C^{(2)} + \mathbf{B^{(2)}}^T \mathbf X^{-1}\mathbf B^{(2)}\right)^{-1}
  \mathbf{B^{(2)}}^T \mathbf X^{-1} \\
 & \downarrow \nonumber
\end{align}

\begin{align}
\ket {x^{(3)}} &= \ket{ \red{i_1} \red{i_2} \red{i_3} {j_4} } \\
\mathbf X^{(3)} &= \left[ \begin{array}{cccc}
        0              &  \blu{F_{i_1}^{i_2}} &  \myi{F_{i_1}^{i_3}} & \red{F_{i_1}^{j_4}} \\
  \blu{-F_{i_1}^{i_2}} &       0              &  \myi{F_{i_2}^{i_3}} & \red{F_{i_2}^{j_4}} \\
  \myi{-F_{i_1}^{i_3}} & \myi{-F_{i_2}^{i_3}} &       0              & \red{F_{i_3}^{j_4}} \\
  \red{-F_{i_1}^{j_4}} & \red{-F_{i_2}^{j_4}} & \red{-F_{i_3}^{j_4}}  &      0
\end{array}\right]
 = \mathbf X + \mathbf B^{(3)} \mathbf C^{(3)} \mathbf{B^{(3)}}^T \\
 \label{eq:pfu_pfa3}
\braket{x^{(3)}}{\phi} &= \PfX^{(3)} = \PfX \times \left.
 \mathrm{Pf} \left(\mathbf C^{(3)} + \mathbf{B^{(3)}}^T \mathbf X^{-1}\mathbf B^{(3)}\right) \right/
  \PfC^{(3)} \\
 \label{eq:pfu_inv3}
\mathbf{X^{(3)}}^{-1} &= \mathbf X^{-1} - \mathbf X^{-1} \mathbf B^{(3)}
 \left(\mathbf C^{(3)} + \mathbf{B^{(3)}}^T \mathbf X^{-1}\mathbf B^{(3)}\right)^{-1}
  \mathbf{B^{(2)}}^T \mathbf X^{-1} \\
 & \vdots \nonumber
\end{align}

Matrices $\mathbf B^{(r)}$ and $\mathbf C^{(r)}$ appearing in \cref{eq:pfu_pfa1,eq:pfu_pfa2,eq:pfu_pfa3,eq:pfu_inv1,eq:pfu_inv2,eq:pfu_inv3} also have a common expression for their entries:
\begin{align}
B^{(r)}_{lm} &= \left\{ \begin{array}{ll}
  F_{i_r}^{j_l} - F_{j_r}^{j_l} & l > m, m \le r \\
  F_{i_r}^{i_l} - F_{j_r}^{i_l} & l < m, m \le r \\
  1                             & l = m, m > r \\
  0                             & \textrm{otherwise},
\end{array}\right. \\
\mathbf C^{(r)}_{lm} &= \begin{bmatrix}
  \mathbf 0 &-\mathbf I_r \\
  \mathbf I_r & \mathbf 0
\end{bmatrix}.
\end{align}
Note also that we have assumed in these equations that electrons in $\ket x$ change their positions by order from the first one. In real implementation of mVMC where the hopping electron is selected randomly, one needs to modify $\mathbf B$ accordingly to select a different set of unit vectors corresponding to that random selection.

\begin{table}[htbp]
    \centering
    \caption{Different ways for evaluating Markov Chain on $\ket x$ with $r_\mathrm{max} = 3$}
\begin{tabular}{l|rrrr}
    \hline
    Configuration $\ket{x^{(k)}}$ &
    \begin{tikzpicture}
        % Node (9) is for creating empty space above.
        \node (9) at (0, 1.3) {};
        \node [style=C1, text width=0.6cm, text height=2.2cm] (4) at (0, 0) {.};%{(0)};%{$\ket{x^{(0)}}$};
        \node [style=Y1] (3) at (0, -0.75) {$j_3$};
        \node [style=Y1] (2) at (0, -0.25) {$j_2$};
        \node [style=Y1] (1) at (0,  0.25) {$j_1$};
        \node [style=Y1] (0) at (0,  0.75) {$j_0$};
        % \node (5) at (1.1, 0) {};
        % \draw[->] (4) to (5);
    \end{tikzpicture} &
    \begin{tikzpicture}
        \node [style=C1, text width=0.6cm, text height=2.2cm] (4) at (0, 0) {.};%{(0)};%{$\ket{x^{(0)}}$};
        \node [style=Y1] (3) at (0, -0.75) {$j_3$};
        \node [style=Y1] (2) at (0, -0.25) {$j_2$};
        \node [style=Y1] (1) at (0,  0.25) {$j_1$};
        \node [style=M1] (0) at (0,  0.75) {$i_0$};
        \node (5) at (-1.5, 0) {$\rightarrow$};
        % \draw[->] (5) to (4);
    \end{tikzpicture} & 
    \begin{tikzpicture}
        \node [style=C1, text width=0.6cm, text height=2.2cm] (4) at (0, 0) {.};%{(0)};%{$\ket{x^{(0)}}$};
        \node [style=Y1] (3) at (0, -0.75) {$j_3$};
        \node [style=Y1] (2) at (0, -0.25) {$j_2$};
        \node [style=M1] (1) at (0,  0.25) {$i_1$};
        \node [style=M1] (0) at (0,  0.75) {$i_0$};
        \node (5) at (-1.5, 0) {$\rightarrow$};
        % \draw[->] (4) to (5);
    \end{tikzpicture} & 
    \begin{tikzpicture}
        \node [style=C1, text width=0.6cm, text height=2.2cm] (4) at (0, 0) {.};%{(0)};%{$\ket{x^{(0)}}$};
        \node [style=Y1] (3) at (0, -0.75) {$j_3$};
        \node [style=M1] (2) at (0, -0.25) {$i_2$};
        \node [style=M1] (1) at (0,  0.25) {$i_1$};
        \node [style=M1] (0) at (0,  0.75) {$i_0$};
        \node (5) at (-1.5, 0) {$\rightarrow$};
        % \draw[->] (4) to (5);
    \end{tikzpicture} \\
    \hline

    Old method evaluates: & --           & \cref{eq:pfu_pfa1,eq:pfu_inv1}
                                         & \cref{eq:pfu_pfa1,eq:pfu_inv1}
                                         & \cref{eq:pfu_pfa1,eq:pfu_inv1} \\
    Old method modifies:  & --           & $\PfX, \InvX$ & $\PfX, \InvX$ & $\PfX, \InvX$ \\
    Old method calls:     & \ttfamily\fontseries{lc}\selectfont SKPFA/SKTDI
                                         & \ttfamily SKMV, SKR2 
                                         & \ttfamily SKMV, SKR2
                                         & \ttfamily SKMV, SKR2 \\
    \hline
    New method evaluates: & --           & \cref{eq:pfu_pfa1}
                                         & \cref{eq:pfu_pfa2}
                                         & \cref{eq:pfu_pfa3,eq:pfu_inv3} \\
    New method modifies:  & --           & $\PfX$        & $\PfX$        & $\PfX, \InvX$ \\
    New method calls:     & \ttfamily\fontseries{lc}\selectfont SKPFA/SKTDI
                                         & \ttfamily SKMV 
                                         & \ttfamily SKMV 
                                         & \ttfamily SKMM, GEMMT \\
    \hline
\end{tabular}
    \label{tab:mc_rank3}
\end{table}

\begin{figure}
    \centering
\begin{tabular}{ll}
\textbf{before}: & \textbf{after}: \\
\begin{tikzpicture}[
             font = \ttfamily\fontseries{lc}\selectfont,
    node distance = 5mm and 15mm,
      start chain = A going below,
      base/.style = {draw, minimum width=20mm, minimum height=8mm,
                     align=center, on chain=A},
 startstop/.style = {base, rectangle, rounded corners, fill=red!30},
   process/.style = {base, rectangle, fill=orange!30},
   hotspot/.style = {base, rectangle, fill=blue!30},
      % io/.style = {base, trapezium,
      %              trapezium left angle=70, trapezium right angle=110,
      %              fill=blue!30},
  decision/.style = {base, diamond, fill=green!30},
  every edge quotes/.style = {auto=right}]
                    ]
  \node [startstop]       {Start w/ $\ket{x^{(0)}}, k \leftarrow 0$};                        % <-- A-1
  \node [process]         {Compute $\PfX$ and $\InvX$};%{\texttt{skpfa} and \texttt{sktdi}}; % <-- A-2
  \node [process]         {Randomly select a new configuration $\ket{x'}$};                  % <-- A-3
  \node [process]         {For $\ket{x'}$ compute $\PfX'$ w/ rank-1 fast-update};            % <-- A-4
  \node [process]         {Compute weight $\braket{x'}{\Psi}$ from $\PfX'$ yielded};         % <-- A-5
  \node [decision]        {Accept?};                                                         % <-- A-6
  \node [process]         {$k \leftarrow k + 1$\\$\ket{x^{(k)}} = \ket{x'}$};                % <-- A-7
  \node [hotspot]         {Rank-1 update $\left (\matX^{(k)} \right)^{-1}$};                 % <-- A-8
  \node [decision]        {$k \ge N_\mathrm{tot}$?};                                         % <-- A-9
  \node [startstop]       {Stop};                                                            % <-- A-10
  \node [process,
         right=1cm of A-6]{$k \leftarrow k + 1$\\$\ket{x^{(k)}} = \ket x$};                  % <-- A-11
  \draw [arrows=-Stealth]
      (A-1)  edge[""]              (A-2)
      (A-2)  edge[""]              (A-3)
      (A-3)  edge[""]              (A-4)
      (A-4)  edge[""]              (A-5)
      (A-5)  edge[""]              (A-6)
      (A-6)  edge["acc."]          (A-7)
      (A-6)  edge["rej."]          (A-11)
      (A-7)  edge[""]              (A-8)
      (A-8)  edge[""]              (A-9)
      (A-9)  edge["yes"]           (A-10)
      ;
  \draw [arrows=-Stealth]
      (A-11) |- ([yshift=-2mm] A-3.east);
  \draw [arrows=-Stealth]
      (A-9)  -- ([xshift=4mm] A-9.east) node [below] {no}
             -- ([xshift=3cm] A-9.east)
             |- ([yshift=2mm] A-3.east);
\end{tikzpicture} &
%
%%%
%%%%%
%%%
%
\begin{tikzpicture}[
             font = \ttfamily\fontseries{lc}\selectfont,
    node distance = 5mm and 15mm,
      start chain = A going below,
      base/.style = {draw, minimum width=20mm, minimum height=8mm,
                     align=center, on chain=A},
 startstop/.style = {base, rectangle, rounded corners, fill=red!30},
   process/.style = {base, rectangle, fill=orange!30},
   hotspot/.style = {base, rectangle, fill=blue!30},
      % io/.style = {base, trapezium,
      %              trapezium left angle=70, trapezium right angle=110,
      %              fill=blue!30},
  decision/.style = {base, diamond, fill=green!30},
  every edge quotes/.style = {auto=right}]
                    ]
  \node [startstop]       {Start w/ $\ket{x^{(0)}}, k \leftarrow 0$};                        % <-- A-1
  \node [process]         {Compute $\PfX$ and $\InvX$};%{\texttt{skpfa} and \texttt{sktdi}}; % <-- A-2
  \node [process]         {$r \leftarrow 1$};                                                % <-- A-3
  \node [process]         {Randomly select a new configuration $\ket{x'}$};                  % <-- A-4
  \node [process]         {For $\ket{x'}$ compute $\PfX'$ w/ rank-$r$ fast-update};          % <-- A-5
  \node [process]         {Compute weight $\braket{x'}{\Psi}$ from $\PfX'$ yielded};         % <-- A-6
  \node [decision]        {Accept?};                                                         % <-- A-7
  \node [process]         {$k \leftarrow k + 1,\; r \leftarrow r + 1$\\$\ket{x^{(k)}} = \ket{x'}$};
  \node [decision]        {$k \ge N_\mathrm{tot}$?};                                         % <-- A-9
  \node [startstop]       {Stop};                                                            % <-- A-10
  \node [hotspot,
         right=6mm of A-3]{Rank-$r$ update $\left( \matX^{(k)} \right)^{-1}$};               % <-- A-11
  \node [decision,
         right=12mm of A-8]{$r < r_\mathrm{max}$?};                                          % <-- A-12
  \node [process,
         right=1cm of A-7]{$k \leftarrow k + 1$\\$\ket{x^{(k)}} = \ket x$};                  % <-- A-13
  \draw [arrows=-Stealth]
      (A-1)  edge[""]              (A-2)
      (A-2)  edge[""]              (A-3)
      (A-3)  edge[""]              (A-4)
      (A-4)  edge[""]              (A-5)
      (A-5)  edge[""]              (A-6)
      (A-6)  edge[""]              (A-7)
      (A-7)  edge["acc."]          (A-8)
      (A-7)  edge["rej."]          (A-13)
      (A-8)  edge[""]              (A-9)
      (A-9)  edge["yes"]           (A-10)
      (A-11) edge[""]              (A-3)
      ;
  \draw [arrows=-Stealth]
      (A-13) |- ([yshift=-2mm] A-4.east);
  \draw [arrows=-Stealth]
      (A-12) -- ([xshift=1mm] A-12.east) node [below] {no}
             |- (A-11);
  \draw [arrows=-Stealth]
      (A-12) -- (A-12.north) node [above right] {yes}
             |- ([xshift=1cm, yshift=2mm] A-4.east)
             -- ([yshift=2mm] A-4.east);
  \draw [arrows=-Stealth]
      (A-9)  -- (A-9.east) node [below right] {no}
             -| (A-12);
\end{tikzpicture}
\end{tabular}
    \caption{Flowcharts for evaluating Markov chain on $\ket x$ \textbf{before} and \textbf{after} block-update optimization.}
    \label{fig:mc_flowchart}
\end{figure}
\paragraph{Application}
With \cref{eq:pfu_pfa1,eq:pfu_pfa2,eq:pfu_pfa3,eq:pfu_inv1,eq:pfu_inv2,eq:pfu_inv3}, mVMC is able to proceed throughout the MC without recalculating $\PfX$ upon each sample. However, in the current version of mVMC, only rank-1 update equations (i.e. \cref{eq:pfu_pfa1,eq:pfu_inv1}) are used which eventually made the program memory-bound.
To get around this problem to make full use of modern CPU architectures, we extend the way of evaluating MC on $\ket{x^{(k)}}$ employed in mVMC. In our new approach:
\begin{itemize}
    \item Rank-$r |_{r \in \mathbb Z^+}$ (blocked) updates are used instead of just rank-1 updates;
    \item A tuning parameter $r_\mathrm{max}$ is set to determine at which step should $\matX^{-1}$ be updated. Note that rank-$r$ update does \emph{not} require $\left(\matX^{(l)} \right)^{-1}$ from previous steps.
\end{itemize}
\Cref{tab:mc_rank3} gives an example of how this change looks like when operating on a 4-electron configuration with $r_\mathrm{max} = 3$ and \cref{fig:mc_flowchart} compares the flowcharts before and after our modification. It is important to note that our optimization does not alter sampling process or acceptance ratio of the original Markov chain. What we have done is to ``pack up'' the heaviest computation, i.e. update of $\InvX$, grouping each $r_\mathrm{max}$ rank-1 updates into one rank-$r_\mathrm{max}$ update so that the new program becomes much faster as we will see in \cref{sec:benchmark}.

\section{Library Interface and Accelerated Implementation}
\label{sec:api}
This work is basically an optimization of the open-source program mVMC. We did not make any changes on the user side interface defined in \cite{mVMC2019}. Instead, this section gives a brief specification of library routines we have made to implement procedures presented in \cref{sec:pfaffine,sec:pfupdates}.

\begin{figure}[htbp]
    \centering
    \begin{tikzpicture}
		\node [style=Y1] (0) at (-1, 1) {Pfaffine};
		\node [style=Y1] (1) at (1, 1) {BLIS};
		\node [style=C1,align=left]
		                          (2) at (-1, -0.35) {\texttt{skpfa}\\\texttt{sktdf}\\\texttt{sktdi}};
		\node [style=C1,align=left]
		                          (3) at (1, -0.05) {\texttt{skr2k}\\\texttt{skmm}};
        \node [style=Y1] (20) at (-1, -1.75) {PfUpdates};
		\node [style=C1,align=left]
		                          (21) at (-1, -3.1) {\texttt{push()}\\\texttt{pop()}\\\texttt{merge()}};

		\node (4) at (-1.1, 1.75) {LAPACK Lv.};
		\node (5) at ( 1.0, 1.75) {BLAS Lv.};
		\node (10) at (0.1, 2) {};
		\node (11) at (0.1, -3.5) {};
		\node [style=M1,align=left] 
		                          (6) at (4, -0.05)
		    {Skew-symmetric\\
		     Assembly: A64FX\\
		     Assembly: Skylake-X\\
		     Assembly: Zen, etc.};
		\draw     (0) to (2);
		\draw[->] (0) to (1);
		\draw     (1) to (3);
		\draw[->] (3) to (6);
		\draw     (20) to (21);
		\draw[dotted] (10) to (11);
        \draw
            (20)   -- ([xshift=2mm] 20.east)
                   |- (1);
    \end{tikzpicture}
    \caption{Two-layer structured solution for computing Pfaffian and inverse of skew-symmetric matrix $\matX$.}
    \label{fig:pfaffine}
\end{figure}
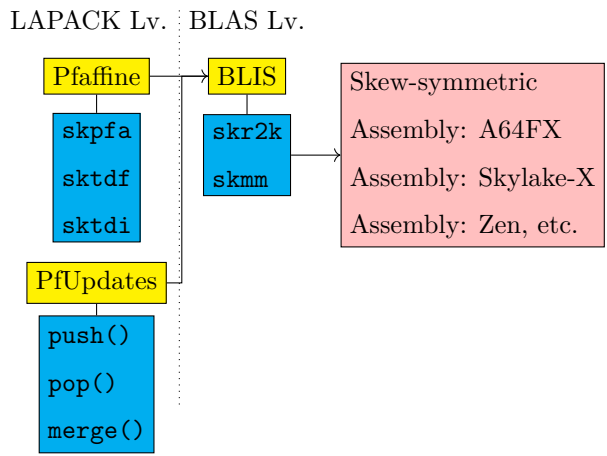

\paragraph{Layered Implementation}
As was stated in \cref{sec:intro}, Wimmer's algorithm is good but their implementation Pfapack77 is not tuned to use assembly-level optimized BLAS routines as skew-symmetric operation is not part of the standard. To complete a high-performance implementation for evaluating equations appeared in \cref{sec:pfaffine,sec:pfupdates}, an extended BLAS-level implementation is needed.
BLIS\cite{BLIS1} is a fast and highly portable framework that implements BLAS-like operations with a fresh approach based on the GotoBLAS\cite{Goto2,Goto1} algorithm. It is able to instantiate multiple level-3 BLAS operations with a minimal number of assembly-based kernels\cite{BLIS2} tuning based on an analytical model\cite{BLIS4}. By adding a minimum amount of interface code to BLIS, skew-symmetric BLAS operations are imported with optimized kernels for AMD Zen, Intel Skylake-X and many other architectures, among them a set of \texttt{GEMM} kernels designed for A64FX\footnote{
    A64FX is the processor name of current world top supercomputer ``Fugaku''.
    } made by ourselves in another joint work with Forschungszentrum Jülich.
Integrating these modules we have now built a layered approach shown in \cref{fig:pfaffine} for computing and block-updating Pfaffian which is fast, well-organized and runs across different CPU architectures.

\begin{table}[htbp]
    \centering
    \caption{List of API handles of the Pfaffine library for Pfaffian and inverse computation. Here $T$ refers to different floating-point data types and $t$ their LAPACK-style prefix characters. We use ``sk. matrix'' as the shorthand of a skew-symmetric matrix.}
    \begin{tabular}{lllll}
        \hline
        \textbf{C++} & \textbf{C (Simp.)} & \textbf{C/Fort.} & \textbf{Description} \\
        \hline
        \texttt{sktdf<}$T$\texttt{>} & N/A & N/A & Tridiagonal factorize $\mathbf X$ (\cref{eq:sktdf}). \\
        \texttt{sktdi<}$T$\texttt{>} & N/A & N/A & Inverse matrix from Tridiagonal factorization (\cref{eq:sktdi}). \\
        \texttt{skpfa<}$T$\texttt{>} & $t$\texttt{skpfa\_} & \texttt{cal\_}$t$\texttt{skpfa\_} 
                                                 & Tridiagonal factorize and compute $\PfX$ (\cref{eq:skpfa}), \\
                                     &     &     & $\mathbf X^{-1}$ computed conditionally depending on input. \\
        \hline
        \texttt{skslc<}$T$\texttt{>} & N/A & N/A & Get a column from corner-stored sk. matrix. \\
        \texttt{skswp<}$T$\texttt{>} & N/A & N/A & Swap columns and rows of cornet-stored sk. matrix. \\
        \texttt{skr2k<}$T$\texttt{>} & N/A & N/A & $\mathbf X + \mathbf{AB}^T - \mathbf{BA}^T$
                                                   where $\mathbf X$ is corner-stored sk. matrix. \\
        \texttt{ skmm<}$T$\texttt{>} & N/A & N/A & $\mathrm{XB}$
                                                   where $\mathbf X$ is corner-stored sk. matrix. \\
        \hline
    \end{tabular}
    \label{tab:api_pfaffine}
\end{table}

\begin{table}[htbp]
    \centering
    \caption{Method API of C++ object \texttt{updated\_tdi} for rank-$r$ update strategy.}
    \begin{tabular}{ll}
        \hline
        \textbf{C++ Method} & \textbf{Description} \\
        \hline
        \texttt{updated\_tdi<}$T$\texttt{>::updated\_tdi()} & Constructor.
                                                              Given $f_{ij}$ and $\ket x$, compute $\PfX$ and $\mathbf X^{-1}$. \\
        \texttt{updated\_tdi<}$T$\texttt{>::get\_Pfa()} & Get Pfaffian of current configuration. \\
        \texttt{updated\_tdi<}$T$\texttt{>::push\_update()} & Push a configuration update: $\ket{x^{(r)}} \rightarrow \ket{x^{(r+1)}}$ \\
                                                            & with one electron changed its spin-orbital position. \\
        \texttt{updated\_tdi<}$T$\texttt{>::pop\_update()} & Revert the last \texttt{push\_update()}. \\
        \texttt{updated\_tdi<}$T$\texttt{>::merge\_updates()} & Modify $\mathbf X^{-1}$ based on the last accepted configuration. \\
        \hline
    \end{tabular}
    \label{tab:api_pfupdates}
\end{table}

\paragraph{Interface}
C++ and C interfaces of our Pfaffian-inverse framework described in \cref{fig:pfaffine} are listed in \cref{tab:api_pfaffine}, while object-based routines for \cref{sec:pfupdates}'s blocked update are listed in \cref{tab:api_pfupdates}. We recommend referring to our program distribution in
\cite{blockedvmc} for more API-related information.

\section{Benchmarking Results}
\label{sec:benchmark}
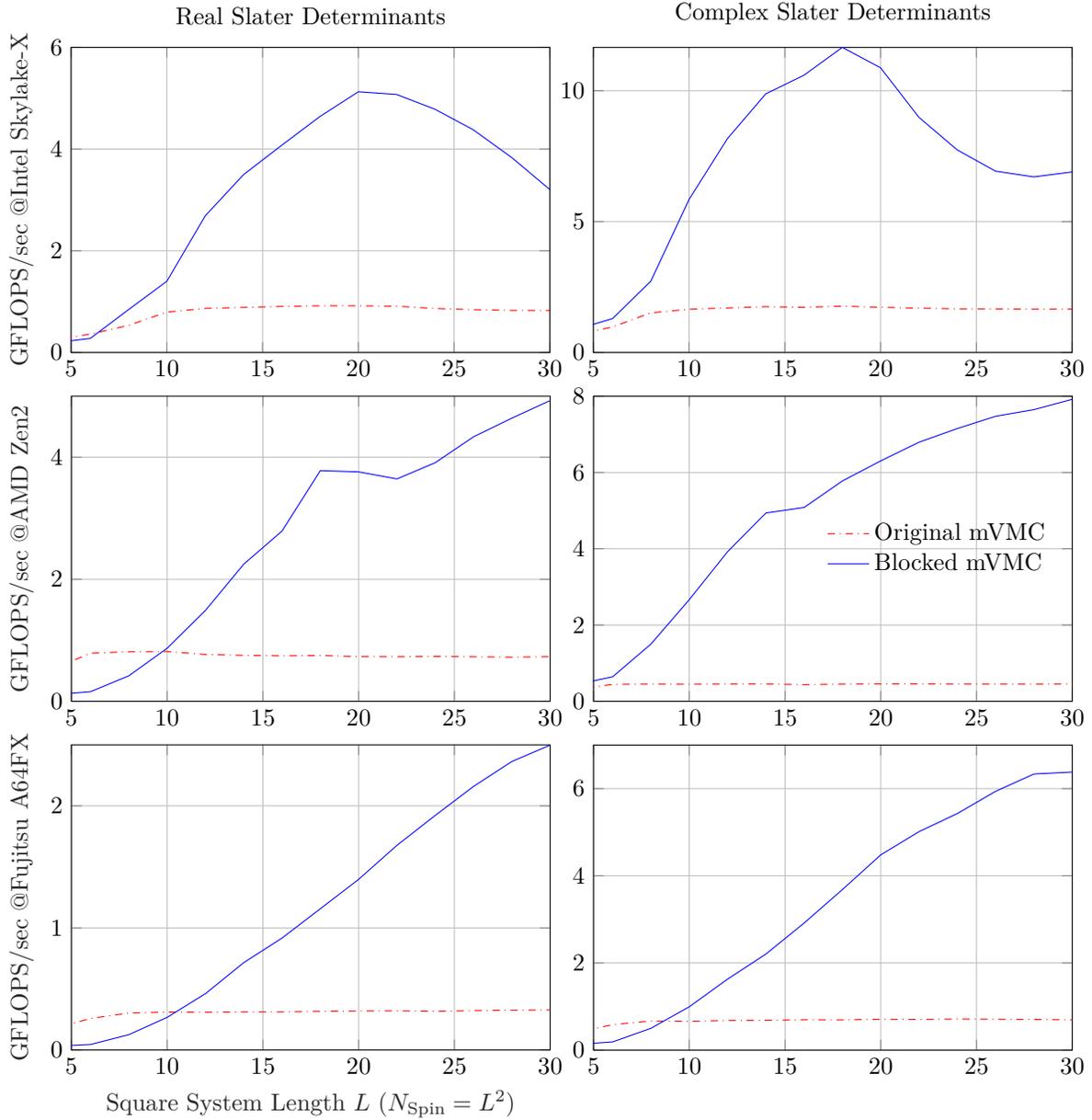
\begin{figure}
    \centering
\begin{tikzpicture}

\begin{axis}[%
width=2.726in,
height=1.738in,
at={(0.248in,4.222in)},
scale only axis,
xmin=5,
xmax=30,
ymin=0,
ymax=6,
ylabel style={font=\color{white!15!black}},
ylabel={GFLOPS/sec @Intel Skylake-X},
axis background/.style={fill=white},
title={Real Slater Determinants},
xmajorgrids,
ymajorgrids
]
\addplot [color=red, dashdotted, forget plot]
  table[row sep=crcr]{%
0 0.0\\
6	0.361360642752065\\
8	0.532296644976695\\
10	0.794942821543725\\
12	0.866825974437258\\
14	0.886842654160376\\
16	0.905629948386682\\
18	0.919198340509686\\
20	0.918083208346844\\
22	0.910220760906638\\
24	0.865769246894286\\
26	0.84237831023014\\
28	0.827578760903765\\
30	0.826516341347516\\
};
\addplot [color=blue, forget plot]
  table[row sep=crcr]{%
0 0.0\\
6	0.276356377378085\\
8	0.843291915702509\\
10	1.40104760150203\\
12	2.68518644973394\\
14	3.49602129483801\\
16	4.07594285425816\\
18	4.64214860260482\\
20	5.12765785734162\\
22	5.07437927213826\\
24	4.78452856858324\\
26	4.38021213065174\\
28	3.83686820761082\\
30	3.20387530044223\\
};
\end{axis}

\begin{axis}[%
width=2.726in,
height=1.738in,
at={(3.221in,4.222in)},
scale only axis,
xmin=5,
xmax=30,
ymin=0,
ymax=11.6572118405419,
axis background/.style={fill=white},
title={Complex Slater Determinants},
xmajorgrids,
ymajorgrids
]
\addplot [color=red, dashdotted, forget plot]
  table[row sep=crcr]{%
0 0.0\\
6	0.975453781248362\\
8	1.51410513229492\\
10	1.65199704882212\\
12	1.70201665388671\\
14	1.75076775214859\\
16	1.72741048762184\\
18	1.76958967389285\\
20	1.7286043427835\\
22	1.69334426314391\\
24	1.66833743293429\\
26	1.66595373655769\\
28	1.65403134148631\\
30	1.65689382397753\\
};
\addplot [color=blue, forget plot]
  table[row sep=crcr]{%
0 0.0\\
6	1.29401174480313\\
8	2.72408246208882\\
10	5.84993413447321\\
12	8.17045354013864\\
14	9.8802472705105\\
16	10.5945783948181\\
18	11.6572118405419\\
20	10.8752065673326\\
22	8.98534834039523\\
24	7.74028955067193\\
26	6.92791077725571\\
28	6.70862394536668\\
30	6.89709072740619\\
};
\end{axis}

\begin{axis}[%
width=2.726in,
height=1.738in,
at={(0.248in,2.235in)},
scale only axis,
xmin=5,
xmax=30,
ymin=0,
ymax=5,
ylabel style={font=\color{white!15!black}},
ylabel={GFLOPS/sec @AMD Zen2},
axis background/.style={fill=white},
xmajorgrids,
ymajorgrids
]
\addplot [color=red, dashdotted, forget plot]
  table[row sep=crcr]{%
0 0.0\\
6	0.787604883462819\\
8	0.812904853596715\\
10	0.815602907306597\\
12	0.766725063729931\\
14	0.753864601313665\\
16	0.744631224152567\\
18	0.750919970500139\\
20	0.7324089327175\\
22	0.729766187745863\\
24	0.735173406011932\\
26	0.729862463883927\\
28	0.723791747045999\\
30	0.731020146286318\\
};
\addplot [color=blue, forget plot]
  table[row sep=crcr]{%
0 0.0\\
6	0.157276595744681\\
8	0.417748555167507\\
10	0.869564603638761\\
12	1.4896011020368\\
14	2.24770328844899\\
16	2.79022926594513\\
18	3.78022322329176\\
20	3.75982891176252\\
22	3.64447622992556\\
24	3.91050102359651\\
26	4.33511760577963\\
28	4.6405084845908\\
30	4.92869155830194\\
};
\end{axis}

\begin{axis}[%
width=2.726in,
height=1.738in,
at={(3.221in,2.235in)},
scale only axis,
xmin=5,
xmax=30,
ymin=0,
ymax=8,
axis background/.style={fill=white},
xmajorgrids,
ymajorgrids,
legend style={at={(0.97,0.5)}, anchor=east, legend cell align=left, align=left, fill=none, draw=none}
]
\addplot [color=red, dashdotted]
  table[row sep=crcr]{%
0 0.0\\
6	0.444645387546435\\
8	0.450947921939973\\
10	0.4475353769087\\
12	0.451880228725421\\
14	0.452502584166316\\
16	0.437727228060083\\
18	0.4499132442055\\
20	0.454230683577054\\
22	0.454318822331535\\
24	0.451102478030378\\
26	0.450342411230345\\
28	0.449857342859122\\
30	0.452905476902044\\
};
\addlegendentry{Original mVMC}

\addplot [color=blue]
  table[row sep=crcr]{%
0 0.0\\
6	0.64298636942015\\
8	1.50184150368957\\
10	2.6652799846782\\
12	3.92267078132881\\
14	4.93735060917295\\
16	5.08304987182869\\
18	5.77918971650319\\
20	6.30501101573384\\
22	6.79520045586756\\
24	7.15350626121068\\
26	7.47444626978662\\
28	7.64888581895958\\
30	7.91921731591751\\
};
\addlegendentry{Blocked mVMC}

\end{axis}

\begin{axis}[%
width=2.726in,
height=1.738in,
at={(0.248in,0.248in)},
scale only axis,
xmin=5,
xmax=30,
xlabel style={font=\color{white!15!black}},
xlabel={Square System Length $L$ ($N_{\text{Spin}} = L^2$)},
ymin=0,
ymax=2.5,
ylabel style={font=\color{white!15!black}},
ylabel={GFLOPS/sec @Fujitsu A64FX},
axis background/.style={fill=white},
xmajorgrids,
ymajorgrids
]
\addplot [color=red, dashdotted, forget plot]
  table[row sep=crcr]{%
0 0.0\\
6	0.257596988437752\\
8	0.3019140722792\\
10	0.310079626535265\\
12	0.308442523392695\\
14	0.31182180217456\\
16	0.311923657944739\\
18	0.315656068636486\\
20	0.319076848777649\\
22	0.320650242540282\\
24	0.316146852192856\\
26	0.321934259412796\\
28	0.325788364229337\\
30	0.327908795273134\\
};
\addplot [color=blue, forget plot]
  table[row sep=crcr]{%
0 0.0\\
6	0.0445231875335491\\
8	0.125053350726732\\
10	0.266749462928537\\
12	0.460940475711656\\
14	0.715411675498654\\
16	0.916753115390395\\
18	1.1569465943857\\
20	1.39770267513074\\
22	1.6762571256027\\
24	1.9215372936486\\
26	2.15929022082482\\
28	2.36358104039274\\
30	2.4980198434941\\
};
\end{axis}

\begin{axis}[%
width=2.726in,
height=1.738in,
at={(3.221in,0.248in)},
scale only axis,
xmin=5,
xmax=30,
ymin=0,
ymax=7,
axis background/.style={fill=white},
xmajorgrids,
ymajorgrids
]
\addplot [color=red, dashdotted, forget plot]
  table[row sep=crcr]{%
0 0.0\\
6	0.584827847398602\\
8	0.66728617929063\\
10	0.660882107533603\\
12	0.679450376807439\\
14	0.683270512547549\\
16	0.696236688236141\\
18	0.692973407901574\\
20	0.709121962516362\\
22	0.704725567845075\\
24	0.713937763976998\\
26	0.709805080916189\\
28	0.703955524270424\\
30	0.693733766332668\\
};
\addplot [color=blue, forget plot]
  table[row sep=crcr]{%
0 0.0\\
6	0.188595682131373\\
8	0.500563913836375\\
10	0.994808592027483\\
12	1.62841042300956\\
14	2.20433410570512\\
16	2.91669399761438\\
18	3.68541708086512\\
20	4.47947988926687\\
22	5.01276664333967\\
24	5.42832792358959\\
26	5.93915915818924\\
28	6.33555941675262\\
30	6.379977014766\\
};
\end{axis}

\begin{axis}[%
width=6.194in,
height=6.208in,
at={(0in,0in)},
scale only axis,
xmin=0,
xmax=1,
ymin=0,
ymax=1,
axis line style={draw=none},
ticks=none,
axis x line*=bottom,
axis y line*=left
]
\end{axis}
\end{tikzpicture}%
    \caption{
        Benchmark conducted on 3 different CPUs against mVMC's Markov chain sampling process comparing original version and our block-update optimized version. FLOPs numbers here, given by \cref{eq:tim2flops}, show how much our method has speed up the mVMC program. Performance draw-back of Intel Skylake-X at $L \ge 20$ is believed to be caused by the block-update's panel size exceeding processor's L2 cache\cite{Goto2002}. Future work on deploying a hierarchy storage (e.g. one implemented in \citet{FLASH:TR}) to $\matX$ might address this problem and allow further speed-up.
    }
    \label{fig:bmk_3machines}
\end{figure}

We run a Markov chain sampling a random many-body spin wavefunction on a $L\times L$ square lattice.
Our experiment only studies the sampling process itself against this random wavefunction so that the Hamiltonian is omitted.
The total number of samples generated by the MC is proportional to the number of spins:
$$
    N_\mathrm{tot} \propto N_\mathrm{spin} = L\times L,
$$
because MC needs more steps to decorrelate as system size grows.
This experimental setup yields:
$$
    r_\mathrm{acc} N_\mathrm{proj} \left(N_\mathrm{qty} + 1 \right) \overbrace{ N_\mathrm{smp} N_\mathrm{spin} }^{N_\mathrm{tot}} \left(2 N_\mathrm{spin}^2 + \frac34 \frac1{N_\mathrm{spin}} N_\mathrm{spin}^3 \right) \equiv \omega N_\mathrm{spin}^3 \text{ MACs}
$$
for each system size. Here $r_\mathrm{acc}$ is the acceptance ratio since rank-1 (rank-$k$) update is requested every time (every $k$ times) a configuration update is \emph{accepted}. $N_\mathrm{qty}$, $N_\mathrm{smp}$ and $N_\mathrm{proj}$ correspond to \texttt{NDataQtySmp}, \texttt{NVMCSample} and \texttt{NQ} parameters in the mVMC application. The extra $\frac 34 \frac1{N_\mathrm{spin}} N_\mathrm{spin}^3$ term inside the bracket accounts for the recalculation of full $\matX \rightarrow \matX^{-1}$ every $N_\mathrm{spin}$ acceptances to remove the accumulated round-up error by fast updates.
% Each initial sample yields $O(N_\mathrm{spin}^3)$ MACs while fast-update is performed $\propto N_\mathrm{spin}$ times requiring $O(N_\mathrm{spin}^2 r / r) = O(N_\mathrm{spin}^2)$ each.
% As a result, we can use the parameter $\left. N_\mathrm{spin}^3 \middle/ T_\mathrm{elapsed} \right.$ to evaluate how well the code performs on a certain CPU.
We set $N_\mathrm{qty} = 10$, $N_\mathrm{smp} = 10$ and $N_\mathrm{proj} = 16$ in our experiment. Acceptance ratios turned out to be $r_\mathrm{acc}^\mathrm{real} \approx 31\%$ for real Slater determinants and $r_\mathrm{acc}^\mathrm{comp} \approx 47\%$ for complex ones regardless of $N_\mathrm{spin}$. Hence the coefficients $\omega$ become $\omega_\text{real}\approx 2050$ and $\omega_\text{complex}\approx 3000$ for real and complex Slater determinants, respectively. Since different hardware vendors sometimes count performance data differently, here we use the expression:
\begin{equation} \label{eq:tim2flops}
    \text{FLOPs/sec} = \left. \omega_\text{data} \times \text{OpCost}_\text{data} \times L^6 \middle/ T_\text{elapsed} \right.
        % = \left\{\begin{array}{ll}
        %     4.1 \times 10^3 & \text{real Slater determinants} \\
        %     2.4 \times 10^4 & \text{complex Slater determinants}
        % \end{array}\right.
\end{equation}
to convert time measurements to floating-point performance data, where $\text{OpCost}_\text{data}$ is the number of floating-point operations (FLOPs) yielded by each MAC on corresponding data types, hence $2$ for real Slater determinants and $8$ for complex Slater determinants.

Three CPU variants are selected to be included in this test, each having its own significance in the HPC context:
\begin{itemize}
    \item Intel\textregistered{} Xeon\textregistered{} Platinum 8260 under Skylake-X microarchitecture with AVX512 vector instructions support;
    \item AMD EPYC\texttrademark{} 7702 under Zen 2 microarchitecture with AVX2 vector instructions support;
    \item Fujitsu A64FX based on Armv8 ISA with 512-bit Scalable Vector Extension\cite{ArmSVE2018} support.
\end{itemize}
Results of our blocked-update mVMC compared to the original implementation is shown in \cref{fig:bmk_3machines}. Note that all tests here are conducted single-threaded because mVMC has parallel capability at a higher level. These results show that performance of our blocked algorithm can ramp up quickly with the growth of system size,
%\footnote{That our algorithm is giving lower performance at small sizes is due to our strategy of selecting $r_\mathrm{max}$ not considering special \texttt{GEMM} microkernels used by small matrices. We are still working on further patches for small matrices but the real research interest is usually oriented to large-size calculations.}
yielding up to more than $6\times$ performance compared to the original code \footnote{The original code is compiled with corresponding vendor compilers under default max-optimization options.} on all our processors tested, while the latter two tested architectures seem to bear potential for higher speed-up rates on even larger system sizes. Among all test cases here, AMD EPYC\texttrademark{} 7702 yields the most significant speed-up (up to $\sim 17.5\times$) thanks to a new skinny-matrix handling strategy introduced to BLIS\cite{blisgemmsup}.

\section{Conclusion}
\label{sec:conclusion}
We have shown in this article a framework for computing and updating Pfaffians that is fast, flexible and runs across different microarchitectures. mVMC optimized with this framework has achieved significant speed up compared to its original release. Theory as well as library implementation in this work can also be applied to other kinds of variational methods that deals with skew-symmetric wavefunctions, i.e. Fermions, like TurboRVB\cite{nakano2020}. On the implementation side, this blocking would also allow potential portings of the related methods onto heterogeneous platforms like a general-purpose GPU machine, although the sequential nature of Markov chains might require additional parallelizations to unleash GPUs' full power.
% {
    Further, our work shows a new possible pattern of open-source communities collaborating with hardware people to deliver an assembly-level optimized code for HPC applications. % }

\section*{Acklowdgements}
A64FX assembly kernels for BLIS used in this work are developed in a joint work with Stepan Nassyr from Forschungszentrum Jülich and the Science of High-Performance Computing (SHPC) group from the University of Texas at Austin.
A64FX test results were collected under ``Program for Promoting Researches on the Supercomputer Fugaku'' by MEXT with the project title ``Basic Science for Emergence and Functionality in Quantum Matter: Innovative Strongly-Correlated Electron Science by Integration of ``Fugaku'' and Frontier Experiments'' (hp200132).
%\begin{CJK}{UTF8}{ipxm}量子物質の創発と機能のための基礎科学 ―「富岳」と最先端実験の密連携による革新的強相関電子科学\end{CJK}
Our development work used also the Isambard 2 UK National Tier-2 HPC Service (\url{http://gw4.ac.uk/isambard/}) operated by GW4 and the UK Met Office, and funded by EPSRC (EP/T022078/1).
RGX acknowledge the Global Science Graduate Course (GSGC) program of the University of Tokyo.
TO and ST acknowledge support by the Endowed Project for Quantum Software Research and Education, the University of Tokyo (\url{https://qsw.phys.s.u-tokyo.ac.jp/}).

\bibliography{mybibfile}

\end{document}